\documentclass[14pt]{article}
\usepackage{amsmath,amsfonts,amssymb,theorem,verbatim,graphics,epsfig}
\def\m{c \;} 
\def\G{\Gamma} 

\newtheorem{lem}{Lemma}[section] 
\def\BEN{\begin{enumerate}}  \def\BI{\begin{itemize}}
\def\EEN{\end{enumerate}}   \def\EI{\end{itemize}}
   \def\sec{\section} \def\nn{\nonumber}

\def\beq{\begin{eqnarray}} \def\eeq{\end{eqnarray}}

\def\eqn#1{\begin{equation}#1\end{equation}}

\def\al*#1{\begin{align*}#1\end{align*}}

\def\ga*#1{\begin{gather*}#1\end{gather*}}

\def\alat*#1#2{\begin{alignat*}{#1}#2\end{alignat*}}
\def\bea{\begin{eqnarray*}}
\def\eea{\end{eqnarray*}}
\def\ml*#1{\begin{multline*}#1\end{multline*}}

 \def\mbf{\mathbf} 
\def\mc{\mathcal}  \def\ovl{\overline}

\def\P{{\mathbb P}} \def\le{\left} \def\ri{\right} \def\i{\infty}
\def\E{{\mathbb E}}   \def\R{{\mathbb R}}
\def\N{{\mathbb N}}

\def\te#1{\mathrm{e}^{#1}}

\def\I{\int}      
  \def\d{\delta}   \def\th{\theta}

  \def\nn{\nonumber} \def\p{\pi}  \def\s{\sigma}
     \def\ps{\psi}
 \def\q{\qquad} \def\D{\Delta}
\def\F{\Phi}   
  \def\td{d}
 
\newtheorem{Thm}{Theorem}
\newtheorem{Cor}{Corollary} \newtheorem{Lemma}{Lemma}
 
{\theorembodyfont{\normalfont}
 
\newtheorem{Rem}{Remark} 
 
}

\newcommand{\proof}{{\it Proof\ }}

\newcommand{\exit}{{\mbox{\, \vspace{3mm}}}
\hfill\mbox{$\square$}}

\begin{document}

\title{Dividend problem with Parisian delay for a spectrally negative L\'evy risk process}

\author{Irmina Czarna%
\footnote{Department of Mathematics, University of Wroc\l
aw, pl. Grunwaldzki 2/4, 50-384 Wroc\l aw, Poland, e-mail:
czarna@math.uni.wroc.pl}\ \hspace{0.3cm} Zbigniew
Palmowski\footnote{Department of Mathematics, University of Wroc\l
aw, pl. Grunwaldzki 2/4, 50-384 Wroc\l aw, Poland, e-mail:
zbigniew.palmowski@gmail.com}}
\maketitle

\begin{center}
\begin{quote}
\begin{small}
{\bf Abstract.} In this paper we consider dividend problem for an insurance company whose risk evolves as a spectrally negative L\'{e}vy process (in the absence
of dividend payments) when Parisian delay is applied. The objective function is given by the cumulative
discounted dividends received until the moment of ruin when so-called barrier strategy is applied. Additionally we will consider two possibilities of delay.
In the first scenario ruin happens when the surplus process stays below zero longer than fixed
amount of time $\zeta>0$. In the second case there is a time lag $d$ between decision of paying dividends and its implementation.
\end{small}
\end{quote}
\end{center}

\sec{Introduction}\label{sec:intro}
In risk theory we usually consider classical Cram\'er-Lundberg risk process:
\begin{eqnarray}\label{CL}
 X_t = x + c t -\sum_{i=1}^{N_t} U_i,
\end{eqnarray}
where $x>0$ denotes an initial reserve and $U_i,(i=1,2,...)$ are i.i.d distributed claims with the distribution function $F$.
The arrival process is a homogeneous Poisson process $N_t$ with intensity $\lambda$.
The premium income is modeled by a constant premium density $c$ and the net profit condition is then $\lambda\E U_1/c<1$.
Lately there has been considered more general setting of a spectrally negative L\'evy process. That is,
$X=\{X_t\}_{t\geq 0}$ is a process with stationary and independent increments having nonpositive jumps.
We will assume that process starts from $X_0=x$ and
later we will use convention $\P(\cdot|X_0=x)=\P_x(\cdot)$ and $\P_0=\P$.
Such process takes into account not only large claims compensated by a steady income at rate $c>0$, but also
small perturbations coming from the Gaussian component and additionally (when $\nu(-\infty,0)=\infty$ for the L\'{e}vy measure $\nu$ of $X$)
compensated countable infinite number of small claims arriving over each finite time horizon.
Working under this class of models, it became apparent that, despite of the diversity of possible probabilistic behaviors it allows to express all results in a unifying manner via the $q$-harmonic scale function $W^{(q)}(x)$ defined via its Laplace transform.
This paper further illustrates this aspect, by unveiling the way the scale functions intervenes in a quite complicated control problem.

The
classic research of the scandinavian school had focused on determining the  "ruin probability" of the
process \eqref{CL} ever becoming negative, under the
assumption that $X$ has positive profits.  Since however in this case the surplus  has
the unrealistic property that it converges to infinity with
probability one, De Finetti \cite{Fin} introduced the dividend
barrier model, in which all surpluses above a given level are
transferred (subject to a discount rate) to a beneficiary, and
raised the question of optimizing this barrier.
Formally, we consider
the risk process controlled by the dividend policy $\pi$ given by
\begin{equation}\label{eq:U}
U^\pi_t = X_t - L^\pi_t,
\end{equation}
where $X_0=x>0$  is an initial reserves and $L^\pi_t$ is an
increasing, adapted and left-continuous process representing the cumulative dividends
paid out by the company up till time $t$.
The optimization objective function is given by the average
cumulative discounted dividends received until the moment of ruin:
\begin{equation}\label{eq:wpi}
v^{\pi}(x) = \E_x \int_0^{\sigma^{\pi}}e^{-qt}\td L^{\pi}_t,
\end{equation}
where $\sigma^\pi$ is a ruin time that we specify later depending on the considered scenario
and $q$ is a discounting rate.

The objective of beneficiaries of an insurance company
is to maximize $v^\p(x)$ over all admissible strategies $\p$:
\begin{equation}\label{optdiv}
v_*(x) = \sup_{\pi\in\Pi} v^\pi(x),
\end{equation}
where $\Pi$ is a set of all admissible strategies, that is strategies
$\pi=\{L^\pi_t, t\geq 0\}$
such that $L^\pi_t-L^\pi_{t-}<U_{t-}^\pi$.

For classical risk process (\ref{CL}) an intricate
"bands strategy" solution was discovered by Gerber \cite{Ger69, Ger72}, as well as the fact that for exponential claims, this reduces to a
simple barrier strategy: "pay all you can above a fixed constant
barrier $a$".

There has been  a great deal of work on De Finetti's
objective, usually concerning barrier  strategies. Gerber and Shiu
\cite{GerberEllias} and Jeanblanc and Shiryaev \cite{JeanShir}
consider the optimal dividend problem in a Brownian setting.
Irb\"{a}ck \cite{Irback} and Zhou \cite{Zhou} study the constant
barrier under the Cram\'{e}r-Lundberg model (\ref{CL}).
Hallin \cite{Hal} formulated time dependent
integro-differential equations describing the payoff associated to
a $2 n$ bands policy. The optimality of the "bands strategy" was
recently established by Albrecher and Thonhauser \cite{AT08}
in the presence of fixed interest rates  as well.
For related work considering both excess-of-loss
reinsurance and dividend distribution policies (e.g. in a diffusion
setting), see Asmussen et al. \cite{AHT} and \cite{Paulsen} and references included in this papers, and  for work including
also a utility function, see Grandits et al. \cite{GHS}.

For L\'evy risk process considered in this paper without Parisian delay
Avram et al. \cite {APP}, Loeffen \cite{Loeffen1} and Loeffen and Renaud \cite{renloef}
found sufficient conditions for which barrier strategy is optimal.
In fact Avram et al. \cite{APP2} prove that in this case the bands strategy is optimal.

In this paper we want to analyze the dividend problem when so-called Parisian delay either at moment of dividends payments
or at the ruin time is applied. The name for this delay comes from Parisian option that price
are activated or canceled depending on the type of option if the
underlying asset stays above or below the barrier long enough in a row
(see \cite{Albrechernew} and \cite{DassWu3b}).

For the classical risk process (\ref{CL}) Dassios and Wu \cite{DassWu3} consider a Parisian type delay between a decision to pay a dividend and its implementation.
The decision to pay is taken when the surplus reaches the fixed barrier $a$ but it is implemented only when the surplus stays above barrier longer than fixed
$d>0$. The dividend is paid at the end of this period. This strategy we will denote by $\pi_{a,d}$.
Similar problem for a spectrally negative L\'{e}vy process of bounded variation was analyzed in \cite{LRZ}.
In this paper we generalize this result into general spectrally negative L\'{e}vy risk process.
In this case ruin time is given by: $\sigma^{\pi_{a,d}}=\inf\{t\geq 0: U^{\pi_{a,d}}_t<0\}$.
Since the ruin time is classical we know that band strategy is optimal
and we also know the necessary conditions under which an optimal strategy is the barrier strategy. We still believe
that this new Parisian strategy (although not optimal within all strategies)
could be very useful for the insurance companies giving possibility of natural delay between decision and its implementation.

In this paper we also consider Parisian delay at the ruin. We denote this strategy by $\pi^{a,\zeta}$.
That is ruin occurs if process $U^{\pi^{a,\zeta}}$ stays below zero for longer period than a fixed $\zeta>0$.
Formally, we define Parisian time of ruin by:
\begin{equation}\label{parruin}\sigma^{\pi^{a,\zeta}}=\inf\{t>0: t-\sup\{s\leq t: U^{\pi^{a,\zeta}}_s\geq 0\} \geq \zeta, U^{\pi^{a,\zeta}}_t<0 \}.\end{equation}
We first analyze the strategy $\pi^{a,\zeta}$ for which the
dividends are paid according to classical barrier dividend strategy transferring all surpluses above a
given level $a$ to dividends. We also prove the verification theorem for this type of ruin.
In particular we find sufficient condition for the barrier strategy to be optimal.

In fact combination of both scenarios is also available.
To simplify analysis we decided to skip this possibility.

We believe that giving possibility of Parisian delay could describe better many situations of insurance company.
For example it can be checked if indeed company's reserves increase and we can pay dividends (in the first scenario)
or it can be given possibility for the insurance company to get solvency (in the second scenario).

The paper is organized as follows. In Section \ref{sec:prel} we introduce basic notions and notations. In Section
\ref{sec:div1} we find discounted cumulative dividends payments until Parisian ruin time.
In Section \ref{sec:ver} we prove the verification theorem and find necessary conditions for the barrier strategy to be optimal.
In Section \ref{sec:div2}
we analyze the case when there is a time lag between decision to pay dividends and its implementation.
Section \ref{sec:examples} is devoted to examples.

\sec{Preliminaries}\label{sec:prel}

We first review some fluctuation theory of spectrally negative
L\'{e}vy processes and refer the reader for more background to
Kyprianou \cite{Kbook}, Sato \cite{Sato} and Bertoin \cite{bertbook}
and references therein.

In this paper we consider a spectrally negative L\'evy process $X=\{X_t\}_{t\geq 0}$, that is a L\'evy process with the L\'evy measure $\nu$
satisfying $\nu(0,\infty)=0$ (for simplicity we exclude the case of a compound Poisson process with negative jumps).
Since jumps
of a spectrally negative L\'{e}vy process
$X$ are all non-positive, moment generating function
$\E[\te{\th X_t}]$ exists for all $\th\ge0$ and is given by
$\E[\te{\th X_t}] = \te{t\psi(\th)}$ for some function
$\ps(\th)$
which is strictly convex with the
property that $\lim_{\th\to\i}\ps(\th)=+\i$. Moreover, $\psi$
is strictly increasing on $[\F(0),\i)$,
where $\F(0)$ is the largest root of $\ps(\th)=0$. We shall denote
the right-inverse function of $\ps$ by $\F:[0,\i)\to[\F(0),\i)$.
We will consider also the dual process $\widehat{X}_t=-X_t$ which is a spectrally positive L\'evy process with the L\'{e}vy measure $\widehat{\nu}\left(0,y\right)=\nu\left(-y,0\right)$.
Characteristics of $\widehat{X}$ will be indicated by
using a hat over the existing notation for characteristics of $X$.

For any $\theta$ for which $\ps(\theta) = \log \E[\exp \theta
X_1]$ is finite we denote by $\P^\theta$ an exponential tilting of
measure $\P$ with Radon-Nikodym derivative with respect to
$\P$ given by
\begin{equation}
\label{eq:changeofmeasure} \le.\frac{\td\P^\theta}{\td\P}\ri|_{\mc
F_t} = \exp\le(\theta X_t - \ps(\theta)t\ri),
\end{equation}
where ${\mc
F_t}$ is a right-continuous natural filtration of $X$.
Under the measure $\P^\theta$ the process $X$ is still a
spectrally negative L\'{e}vy process with characteristic function
$\ps_\th$ given by
\begin{equation}\label{tildepsi}
\psi_\theta(s)=\psi(s+\theta)-\psi(\theta).
\end{equation}

Throughout the paper we assume that
the following (regularity) condition is satisfied:
\begin{equation}\label{eq:condW}
\s>0\q\text{or}\q\I_{-1}^0 x \nu(\td x) = \i \q\text{or}\q \nu(\td
x)<<\td x,
\end{equation}
where $\s$ a Gaussian coefficient of $X$.

\subsection{Scale functions}\label{sec:scale}
For $p\geq 0$, there exists a function $W^{(p)}: [0,\i) \to [0,\i)$,
called the {\it $p$-scale function}, that is continuous and
increasing with Laplace transform \eqn{\label{eq:defW} \I_0^\i
\te{-\th x} W^{(p)} (y)  \td y = (\ps(\th) - p)^{-1},\q\q\theta >
\F(p). } The domain of $W^{(p)}$ is extended to the entire real
axis by setting $W^{(p)}(y)=0$ for $y<0$.
We denote $W^{(0)}(z)=W(z)$.
For later use we mention
some properties of the function $W^{(p)}$ that have been obtained
in literature. On $(0,\i)$ the function $z\mapsto W^{(p)}(z)$
is right- and left-differentiable and
under the condition (\ref{eq:condW}), it holds that $z\mapsto
W^{(p)}(y)$ is continuously differentiable for $y>0$. Moreover,
if $\s>0$ it holds that $W^{(p)}\in C^\i(0,\i)$
with $W^{(p)\prime}(0)=2/\s^2$; if $X$ has unbounded
variation with $\s=0$, it holds that $W^{(p)\prime}(0)=\i$
(see \cite[Lemma 4]{P}).

The function $W^{(p)}$ plays a key role in the solution of the
two-sided exit problem as shown by the following classical
identity. Letting $\tau^+_a, \tau^-_a$ be the entrance times of $X$
into $[a,\i)$ and $(-\i,-a)$ respectively,
$$
\tau^+_a = \inf\{t\ge 0: X_t \geq a\},\q \tau^-_a = \inf\{t\ge 0: X_t < -
a\}
$$
it holds for $z\in[0,a]$ that \eqn{\label{twoex} \E_z\le[e^{-p
\tau_a^+},\tau^-_0 > \tau^+_a\ri] = W^{(p)}(z)/W^{(p)}(a). }
Closely
related to $W^{(p)}$ is function $Z^{(p)}$ given by
\begin{equation}\label{Zq}
Z^{(p)}(z) = 1 + q\ovl W^{(p)}(z),
\end{equation}
where $\ovl W^{(p)}(z) = \I_0^zW^{(p)}(y)\td y$ is the
anti-derivative of $W^{(p)}$.
Moreover, the scale functions appear also in so-called
two-sided downward exit problem:
\begin{equation}
 \E_{z}\left[ e^{-p\tau _{0}^{-}},\tau
_{0}^{-}<\tau _{a}^{+}\right] =Z^{(p)}(z)-Z^{(p)}(a)\frac{W^{(p)}(z)%
}{W^{(p)}(a)}.  \label{two-sided-down}
\end{equation}
and in one-sided downward exit problem that
for any  $\beta$ with $\psi(\beta)<\infty$, $p\geq \psi(\beta)\vee 0$ and $x\geq 0$ gives:
\begin{equation}\label{onesideddwon}
\E_z\left[e^{-p\tau_0^- + \beta X_{\tau^-_0}},\tau_0^- <\infty\right]
=e^{\beta z}\left(Z_\beta^{(u)}(z) -\frac{u}{\Phi(u)}W_\beta^{(u)}(z)\right),
\end{equation}
where $W^{(u)}_\beta$ and $Z^{(u)}_\beta$ are scale functions with respect to the measure $\P^\beta$, $u=p-\psi(\beta)$ and
$u/\Phi(u)$ is understood in the limiting sense if $u=0$. In fact for each
$z\in \R$, $W^{(p)}(z)$ is analytically extendable, as a function in $p$, to the whole complex plane; and hence the
same is true of $Z^{(p)}(z)$. In which case arguing again by
analytic extension one may weaken the requirement that $p\geq \psi(\beta)\vee 0$ to simply $p\geq 0$.

The `tilted' scale functions can be
linked to non-tilted scale functions via the
relation $ \te{\beta z} W^{(u)}_\beta(z) =
W^{(p)}(z)$ from \cite[Remark 4]{AKP}. This relation implies that
$$
Z^{(u)}_\beta(z) = 1 + u\I_0^z\te{-\beta y}W^{(p)}(y)\td y.
$$


\subsection{Parisian ruin}

One of most important characteristics in risk theory is a ruin probability defined by
$\P_x(\tau^{-}_0<\infty)$  for $\tau^{-}_0=\inf\{t>0: X_t < 0 \}$. Czarna and Palmowski \cite{iczp}
extended this notion to so-called Parisian ruin probability,
that occurs if the process $X$ stays below zero for period longer than a fixed $\zeta>0$
(see also \cite{DassWu1, DassWu2} for the result concerning classical risk process).
Let
$$\tau^{\zeta}=\inf\{t>0: t-\sup\{s\leq t: X_s\geq 0\} \geq \zeta, X_t>0 \}$$
and Parisian ruin probability we define as:
$$\P(\tau^\zeta<\infty|X_0=x)=\P_x(\tau^\zeta<\infty).$$
The following result summarize \cite[Theorem 1]{iczpr}.

\begin{Thm}\label{ruinprob}
For any $x\geq 0$ Parisian ruin probability equals:
\begin{eqnarray}\nonumber
\label{newformula}
\mathbb{P}_x(\tau^\zeta=\infty) =
 \mathbb E [X_1]\frac{\int_0^\infty   W(x+z)  z\mathbb{P}(X_{\zeta}\in\mathrm{d}z) }{ \int_0^\infty   z\mathbb{P}(X_{\zeta}\in\mathrm{d}z)}.
\end{eqnarray}
\end{Thm}

\sec{Parisian delay at ruin}\label{sec:div1}

In section we will consider Parisian ruin time (\ref{parruin})
and dividends paid according to barrier strategy
that correspond to
reducing the risk process $U^{\pi^{a,\zeta}}$ to the level $a$ if $x>a$, by paying
out the amount $x-a$, and subsequently paying out the minimal
amount of dividends to keep the risk process below the level $a$.
It is well
known (see \cite{APP}) that for $0<x\leq a$ the corresponding controlled risk
process $U^{\pi^{a,\zeta}}$ under $\P_x$ is equal in law to
the process $\{a-Y_t : t\geq 0\}$ under $\P_x$
for \begin{equation}\label{Y}
Y_t=a\vee \overline{X}_t-X_t\end{equation}  being L\'{e}vy process $X$  reflected at its  past supremum:
$$
\overline{X}_t =
\sup_{0\leq s\leq t}X_s,
$$
where we use notations $y\vee0=\max\{y,0\}$.
In this case for all $x\geq 0$,
\[
v^{a,\zeta}(x):=v^{\pi^{a,\zeta}}(x)= \E_x\left( \int_0^{\sigma^{\pi^{a,\zeta}}} e^{-qt}
dL^{\pi^{a,\zeta}}_t\right),
\]
and $L_t^{\pi^{a,\zeta}}
=a\vee\overline{X}_t - a$.

Note that for $x\leq a$,
\begin{equation}\label{basicidentity}
v^{a,\zeta}(x)=\E_x\left[e^{-q\tau_a^+}, \tau_a^+<\tau^\zeta\right]v^{a,\zeta}(a)
\end{equation}
and
\begin{equation}\label{basicidentity2}
v^{a,\zeta}(x)=x-a+v^{a,\zeta}(a)\qquad \mbox{for $x>a$.}
\end{equation}

Assume that $X\to\infty$ a.s. Then by Markov property and fact that
$X$ jumps only downwards we derive:
\begin{equation}\label{twosided1}
\P_x(\tau^\zeta=\infty)=\P_x(\tau_a^+<\tau^\zeta)\P_a(\tau^\zeta=\infty).
\end{equation}
Hence
$$\P_x(\tau_a^+<\tau^\zeta)=\frac{\P_x(\tau^\zeta=\infty)}{\P_a(\tau^\zeta=\infty)}.$$
Using change of measure (\ref{eq:changeofmeasure}) with $\theta=\Phi(q)$, Optional Stopping Theorem
and fact that on $\P^{\Phi(q)}$ process $X$ tends to infinity a.s. (since $\psi_{\Phi(q)}^\prime(0+)=\psi^\prime(\Phi(q)+)>0$), we have for $x\leq a$,
\begin{equation}\label{parisiantwosided}
\E_x\left[e^{-q\tau^+_a},\tau_a^+<\tau^\zeta\right]=\frac{V^{(q)}(x)}{V^{(q)}(a)},
\end{equation}
where by Theorem \ref{ruinprob},
\begin{equation}\label{parisianscale}
V^{(q)}(x)=e^{\Phi(q)x}\P^{\Phi(q)}_x(\tau^\zeta=\infty)=\mathbb E^{\Phi(q)}[ X_1]\frac{\int_0^\infty  e^{-\Phi(q)z} W^{(q)}(x+z)  z\mathbb{P}^{\Phi(q)}(X_{\zeta}\in\mathrm{d}z) }{ \int_0^\infty   z\mathbb{P}^{\Phi(q)}(X_{\zeta}\in\mathrm{d}z)}.\end{equation}

It follows from Theorem \ref{ruinprob} that under the condition (\ref{eq:condW}) function $V^{(q)}(y)$ (similarly like
$W^{(q)}(y)$) is continuously differentiable for $y\in\R$.

Moreover, for $n\in \N$, by (\ref{basicidentity2}),
\begin{eqnarray*}v^{a,\zeta}(a)&\geq& \E_a\left[e^{-q\tau_{a+1/n}^+}, \tau_{a+1/n}^+<\tau^\zeta\right]v^{a,\zeta}\left(a+\frac{1}{n}\right)\\&=&\E_a\left[e^{-q\tau_{a+1/n}^+}, \tau_{a+1/n}^+<\tau^\zeta\right]\left(v^{a,\zeta}(a)+\frac{1}{n}\right)\end{eqnarray*}
and
\begin{eqnarray*}v^{a,\zeta}(a)&\leq& \E_a\left[e^{-q\tau_{a+1/n}^+}, \tau_{a+1/n}^+<\tau^\zeta\right]\left(v^{a,\zeta}(a)+\frac{1}{n}\right)\\&&
+\frac{1}{n}\E_a\left[\int_0^{\tau_{a+1/n}^+}e^{-qt}dt, \tau_{a+1/n}^+<\tau^\zeta\right]\\
&&=\E_a\left[e^{-q\tau_{a+1/n}^+}, \tau_{a+1/n}^+<\tau^\zeta\right]\left(v^{a,\zeta}(a)+\frac{1}{n}\right)\\&&
+\frac{1}{nq}\left(1-\E_a\left[e^{-q\tau_{a+1/n}^+}, \tau_{a+1/n}^+<\tau^\zeta\right]\right)
\end{eqnarray*}
since $L_t^{\pi^{a,\zeta}}=\overline{X}_t-a$ under $\P_a$ can increase only by $\epsilon$ up to time $\tau_{a+1/n}^+$.
Last increment in above equation is ${\rm o}(1/n)$ since by strictly positive drift $X$ is regular for $(0,\infty)$.
Hence,
$$v^{a,\zeta}(a)= \frac{V^{(q)}(a)}{V^{(q)}\left(a+\frac{1}{n}\right)}\left(v^{a,\zeta}(a)+\frac{1}{n}\right)+{\rm o}\left(\frac{1}{n}\right)$$
and
then
$$v^{a,\zeta}(a)=\frac{V^{(q)}(a)}{V^{(q)\prime}(a)}.$$
Thus from (\ref{basicidentity}), (\ref{basicidentity2}) and (\ref{parisiantwosided}) it follows that $v^{a,\zeta}$ is continuously differentiable for all $x\in \R$
and
\begin{eqnarray}
\label{vax} v^{a,\zeta}(x) &=& v^{\pi^{a,\zeta}}(x) =
\begin{cases} \frac{V^{(q)}(x)}{V^{(q)\prime}(a)},
& x \leq a,\\
&\\
x - a + \frac{V^{(q)}(a)}{V^{(q)\prime}(a)}, & x >  a.
\end{cases}
\end{eqnarray}
In particular,
\begin{equation}\label{cosnowego}
(v^{a,\zeta})^\prime(a)=1.
\end{equation}
Hence we get the following theorem.

\begin{Thm}\label{ThmMain1}
The value function corresponding to the barrier strategy $\pi^{a,\zeta}$ is given by
(\ref{vax}).
The optimal barrier $a^*$ is given by:
\begin{equation}\label{cstar}
a^* = \inf\{a>0: V^{(q)\prime}(a) \leq V^{(q)\prime}(y)\ \
\text{for all $y\geq 0$}\}.
\end{equation}
In particular, if $V^{(q)}\in \mathcal{C}^2(\R)$ and there exists unique solution of equation:
\begin{equation}\label{ceq}
V^{(q)\prime\prime}(a^*)=0,
\end{equation}
then $a^*$ is optimal barrier.
\end{Thm}

\begin{Rem}
Note that $V^{(q)}\in \mathcal{C}^2(\R)$ if $W^{(q)}\in \mathcal{C}^2(\R)$. This is the case if e.g. the Gaussian component is present.
\end{Rem}

\sec{Verification Theorem}\label{sec:ver}
To prove the optimality of a particular strategy $\pi$ across all
admissible strategies $\Pi$ for the dividend problem
(\ref{optdiv}), where the ruin time $\sigma^\pi$ is given by the Parisian ruin (\ref{parruin}),
we are led, by standard Markovian arguments, to
consider the following variational inequalities:
\begin{eqnarray}
\Gamma f(x) - q f(x) &\leq& 0,\q \text{if}\quad x \in \R,\label{HJB}\\
f'(x) &\ge& 1,\q \text{if}\quad x\in \R,\label{HJBb}
\end{eqnarray}
for functions $f:\R\to\R$ in the domain of the extended generator
$\Gamma$ of the process $X$ which acts
on $C^2(0,\i)$ functions $f$ as
\begin{equation}\label{Gamma}
\Gamma f(x) = \frac{\s^2}{2}f''(x) + p_0 f'(x) +
\I_{-\infty}^{0}\le[f(x+y)-f(x) + f'(x)y1_{\{|y|<1\}}\ri]\nu(\td y),
\end{equation}
where $\nu$ is the L\'{e}vy measure of $X$ and $\s^2$ denotes the
Gaussian coefficient and $p_0=\m  - \I_{-1}^0y\nu(\td y)$ if the
jump-part has bounded variation; see Sato \cite[Ch. 6, Thm. 31.5]{Sato}. In
particular, if $E|X|<\i$ and $X$ has unbounded variation
[bounded variation], a function $f$ that is $C^2$ $[C^1]$ on
$[0,\infty)$ and that is ultimately linear lies in the domain of
the extended generator.

\begin{Thm}\label{varopt}
Let $C\in(0,\i]$ and suppose $f$ is  continuous and piecewise
$C^1$ on $(-\infty,C)$ if $X$ has bounded variation and that $f$ is
$C^1$ and piecewise $C^2$ on $(-\infty,C)$ if $X$ has unbounded
variation. Suppose that $f$ satisfies (\ref{HJB}) and
(\ref{HJBb}). Then $f\ge\sup_{\pi\in\Pi_{\leq C}} v^\pi$ for $v^\pi$ defined in (\ref{eq:wpi}) with Parisian ruin time (\ref{parruin}), where
$\Pi_{\leq C}$ is a set of all bounded strategies by $C$. In
particular, if $C=\i$, then $f\ge v_*$.
\end{Thm}

\proof We will follow classical arguments. Let $\pi\in\Pi_{\leq
C}$ be any admissible policy and denote by $L=L^{\pi}$ and $U =
U^{\pi}$ the corresponding cumulative dividend process and risk
process, respectively. By Sato \cite[Ch. 6, Thm. 31.5]{Sato}
function $g(t,x, z)=e^{-qt}f(x)\mathbf{1}_{\{z\leq \zeta\}}$ is in a domain of extended generator
of the three-dimensional Markov process $(t,U^\pi_t, \varsigma^U_t)$, with $\varsigma^U_t
=t-\sup\{s\leq t: U_{t}\geq 0\}$.
Note that finite number of discontinuities
in $f$ and hence also single discontinuity in $\mathbf{1}_{\{z\leq \zeta\}}$ are allowed here.
Hence we are also allowed
to apply It\^{o}'s lemma (e.g. \cite[Thm. 32]{Protter}) if $X$
is of unbounded variation and the change of variable formula (e.g.
\cite[Thm. 31]{Protter}) if $X$ is of bounded variation:
\begin{eqnarray}
\nn \te{-qt}f(U_t)\mathbf{1}_{\{\varsigma^U_t\leq \zeta\}} - f(U_0) &=& J_{f}(t) - \I_0^t\te{-qs}f^\prime(U_{s^-})\td L^c_s\\
&&+ \I_0^t\te{-qs}(\Gamma f - qf)(U_{s-})\td s + M_t,
\label{eq:ITO}
\end{eqnarray}
where $M_t$ is a local martingale with $M_0=0$, $L^c$ is the
pathwise continuous parts of $L$
 and for a function $g$ the process $J_g$ is given by
\begin{equation}\label{eq:sumJ}
J_g(t) = \sum_{s\leq t}\te{-qs}\le[ g(A_s + B_s) - g(A_s)\ri]\mbf
1_{\{B_s\neq 0\}},
\end{equation}
where $A_s = U_{s-}+\D X_s$ with $\D x_s=X_s-x_{s-}$ and $B_s = -\D L_s$ denotes the jump
of $-L$ at time $s$. Let $T_n$ be a localizing sequence of $M$.
Applying Optional Stopping Theorem to the stopping times
$T_k^\prime=T_k\wedge \sigma ^\pi$ and using Fatou's Theorem  we
derive:
\begin{eqnarray*}
f(x)&\geq& \E_x\te{-qT_n^\prime}f(U_{T^\prime_n})\mathbf{1}_{\{\varsigma^U_{T^\prime_n}\leq \zeta\}}-J_f(T^\prime_n)+
\E_x\I_0^{T^\prime_n}\te{-qs}f^\prime(U_{s^-})\td L^c_s
\\\\
&&- \E_x\I_0^{T^\prime_n}\te{-qs}(\Gamma f - qf)(U_{s-})\td s.
\end{eqnarray*}
Invoking the variational inequalities $f^\prime(x)\geq 1$ (hence $f(A_s + B_s) - f(A_s)\leq - \Delta L_s$ if
$A_s>0$) and $\Gamma f(x)-qf(x)\leq 0$ we have:
\begin{eqnarray*}f(x)&\geq& \E_x\te{-qT_n^\prime}f(U_{T^\prime_n})\mathbf{1}_{\{\varsigma^U_{T^\prime_n}\leq \zeta\}}
+\E_x\I_0^{T^\prime_n}\te{-qs}\td L_s\\&\geq&
\E_x\left[\te{-q\sigma^\pi}f(U_{\sigma^\pi}); \sigma^\pi\leq
T_n^\prime\right] +\E_x\left[\I_0^{\sigma^\pi}\te{-qs}\td L_s;
\sigma^\pi\leq T_n^\prime\right].
\end{eqnarray*}
Letting $n\to\i$ in conjunction with the monotone convergence
theorem and using fact that $\mathbf{1}_{\{\varsigma^U_{\sigma^{\pi}}\leq \zeta\}}=0$ complete the proof. \exit

Using verification theorem we find necessary conditions under which the optimal
strategy takes the form of a barrier strategy.

\begin{Thm}\label{thm:p1}
Assume that $\s>0$ or that $X$ has bounded variation or, otherwise,
suppose that $v^{a^*,\zeta}\in C^2(0,\i)$.
If $q>0$, then $a^*<\i$ and the following hold true:

(i) $\pi^{a^*, \zeta}$ is the optimal strategy in the set $\Pi_{\leq a^*}$
of all bounded strategies by $a^*$
and $v^{a^*,\zeta} = \sup_{\pi\in\Pi_{\leq a^*}} v^\pi$.

(ii) If $(\G v^{a^*,\zeta} - qv^{a^*,\zeta})(x)\leq 0$ for $x>a^*$, the value
function and optimal strategy of (\ref{optdiv}) is given by $v_*
= v^{a^*,\zeta}$, where the ruin time $\sigma^\pi$ is given by the Parisian moment of ruin (\ref{parruin}).
\end{Thm}

The proof of Theorem \ref{thm:p1} is based on the verification Theorem \ref{varopt} and the following lemma.

\begin{Lemma}\label{fin}
(i) We have $a^*<\infty$.

(ii) It holds that $(\G v^{a^*,\zeta}-qv^{a^*,\zeta})(x)= 0$ for $x\leq a^*$.

(iii) For $x\leq a^*$
$$
(v^{a^*,\zeta})^\prime(x) \geq 1.
$$
\end{Lemma}

\proof
Point (i) follows from the fact that $V^{(q)\prime}(y)$ is continuous and increasing starting at some point. Indeed, note that by \cite{KypPalm}
we have $V^{(q)}(y)=e^{\Phi(q)y}\P_y(\tau^\zeta=\infty)\geq e^{\Phi(q)y}\P_y(\tau_0^-=\infty)=\frac{1}{\psi^\prime(0+)}W^{(q)}(y)$
and $W^{(q)\prime}(y)$ tends to $\infty$ as $y\to\infty$.
The proof of (ii) follows from
(\ref{basicidentity}) and the martingale property of
$$e^{-qt}\E_{X_t}\left[e^{-q\tau_{a^*}^{+}}, \tau_{a^*}^{+}<\tau^\zeta\right]
=\E\left[\E_{x}\left[e^{-q\tau_{a^*}^{+}}, \tau_{a^*}^{+}<\tau^\zeta\right]|{\mc
F_t}\right],$$ where $x\leq a^*$.
Point (iii) is a consequence of (\ref{cosnowego}) and definition of $a^*$ given in (\ref{cstar}).
\exit

Moreover, we can give other necessary condition for the barrier strategy to be optimal.
\begin{Cor}\label{wniosek}
Suppose that
$$V^{(q)\prime}(a)\leq V^{(q)\prime}(b),\qquad\mbox{for all $a^*\leq a\leq b$.}
$$
Then the barrier strategy at $a^*$ is an optimal strategy.
\end{Cor}
\proof
Using Theorems \ref{ThmMain1} and \ref{thm:p1} the proof is the same as the proof of \cite[Theorem 2]{Loeffen1}.\
\exit

\begin{Cor}\label{cor:explopt}
Suppose that, for $x>0$, $\widehat{\nu}'(x)$
is monotone decreasing, then $\pi^{a^*, \zeta}$ is an optimal
strategy of (\ref{optdiv}).
\end{Cor}

\proof
By (\ref{parisianscale}) the proof is similar like the proof of \cite[Theorem 3]{Loeffen1}.
In fact it suffices to prove that $V_{\Phi(q)}$ has completely monotone derivative.
This fact follows from Theorem \ref{ruinprob} since
$\frac{\partial}{\partial x}\P^{\Phi(q)}_x(\tau_0^-=\infty)$ and $\frac{\partial}{\partial x} \P^{\Phi(q)}_x(\tau^{-}_0<\infty, -X_{\tau^{-}_0}\in dz)$ are completely monotone.
Indeed,  it is known that
\begin{equation}\label{ruinprobref}
\frac{\partial}{\partial x}\P^{\Phi(q)}_x(\tau_0^-=\infty)=\widehat{\kappa}_{\Phi(q)}(0,0)\widehat{U}_{\Phi(q)}^\prime(0,x),
\end{equation}
where $\widehat{U}_{\Phi(q)}$ is a renewal function of the descending ladder
height process $\widehat{H}_t$ under $\P^{\Phi(q)}$ and
$\widehat{\kappa}_{\Phi(q)}(\alpha,\beta)$ is a Laplace exponent of bivariate
descending ladder height process $(\widehat{L}_t^{-1}, \widehat{H}_t)$ under $\P^{\Phi(q)}$
with $\widehat{\kappa}_{\Phi(q)}(0,0)=\psi^\prime(\Phi(q))>0$.
From the proof of \cite[Theorem 3]{Loeffen1} it follows that
$\widehat{U}_{\Phi(q)}^\prime(0,x)$ hence also $\frac{\partial}{\partial x}\P^{\Phi(q)}_x(\tau_0^-=\infty)$ is completely monotone.
Moreover, by \cite[(7.15), p. 195]{Kbook} we have
\begin{eqnarray*}\lefteqn{\P^{\Phi(q)}_x(\tau^{-}_0<\infty, -X_{\tau^{-}_0}\in dz)}\\&&=\widehat{\kappa}_{\Phi(q)}(0,0)\int_0^x\widehat{U}_{\Phi(q)}(x-dy)\int_0^\infty e^{-\Phi(q)(z+v)}\widehat{\nu}(dz+v)\;dv,
\end{eqnarray*}
and hence $\frac{\partial}{\partial x} \P^{\Phi(q)}_x(\tau^{-}_0<\infty, -X_{\tau^{-}_0}\in dz)$
is also completely monotone.
\exit

\sec{Parisian delay at the moment of dividend payments}\label{sec:div2}

In this section we analyze the case when we pay dividends only
when surplus process stay above barrier $a$ longer than a time lag $d>0$. The dividends are paid at the end of
that period and they are paid until regular ruin time
$\sigma^{\pi_{a,d}}=\inf\{t\geq 0: U_t^{\pi_{a,d}}<0\}$. Then by (\ref{twoex}) for $x\in[0,a]$,
\begin{equation}\label{basicid1}
v_{a,d}(x):=v^{\pi_{a,d}}(x)=\E_x\left[ e^{-q\tau_a^+}, \tau_a^+<\tau_0^-\right]v_{a,d}(a)=\frac{W^{(q)}(x)}{W^{(q)}(a)}v_{a,d}(a);
\end{equation}
and by Markov property for $x\geq a$,
\begin{eqnarray}
v_{a,d}(x)&=&e^{-q d}\E_{x-a}\left[(X_d-a+v_{a,d}(a)), \tau_0^->d\right]\nonumber\\&&+v_{a,d}(a)\int_{(0,a)}\E_{a-y}\left[ e^{-q\tau_a^+}, \tau_a^+<\tau_0^-\right]\E_{x-a}\left[e^{-q \tau^-_0},-X_{\tau_0^-}\in \td y, \tau_0^-\leq d\right]\nonumber\\&&+v_{a,d}(a)\E_{x-a}\left[e^{-q \tau^-_0},X_{\tau_0^-}=0, \tau_0^-\leq d\right],\label{basicid2}
\end{eqnarray}
where $\E_x\left[ e^{-q\tau_a^+}, \tau_a^+<\tau_0^-\right]$ is given in (\ref{twoex}).

Double Laplace transform of $\E_{z}\left[e^{-q \tau^-_0}, -X_{\tau_0^-}\in \td y, \tau_0^-\leq s\right]$ for $z\geq 0$ by (\ref{onesideddwon}) equals
\begin{eqnarray}
&&\int_0^\infty\int_{[0,\infty)} e^{-\alpha s}e^{-\beta y}\E_{z}\left[e^{-q \tau^-_0}, -X_{\tau_0^-}\in \td y, \tau_0^-\leq s\right]\td s\nonumber\\&&=
\frac{1}{\alpha}\E_{z}\left[e^{-(\alpha+q) \tau^-_0 +\beta X_{\tau_0^-}}, \tau_0^-<\infty\right]\nonumber\\&&=
\frac{1}{\alpha}e^{\beta z}\left(Z_\beta^{(u_q)}(z) -\frac{u_q}{\Phi(u_q)}W_\beta^{(u_q)}(z)\right):=H_q(\beta,z),
\label{trlapid3}
\end{eqnarray}
where $u_q=\alpha+q-\psi(\beta)$. Moreover,
\begin{eqnarray}
&&\int_0^\infty e^{-\alpha s}\E_{z}\left[X_s, \tau_0^->s\right]\td s =
\frac{1}{\alpha}\left\{z-
\E_{z}\left[X_{\tau_0^-}e^{-\alpha\tau^-_0}, \tau_0^-<\infty\right]\right\}\nonumber\\&&=
\frac{1}{\alpha}z-\frac{\partial}{\partial \beta}H_0(\beta,z)_{|\beta=0}.
\label{trlapid4}
\end{eqnarray}

Further, the value $v_{a,d}(a)$ is determined by
(\ref{basicid2}) if $X$ has no Gaussian component ($\sigma =0$) or
by the smooth paste condition:
\begin{equation}\label{smoothpaste}
v_{a,d}^\prime(a-)=v_{a,d}^\prime(a+),
\end{equation}
otherwise.
\begin{lem}
If $\sigma >0$ then (\ref{smoothpaste}) holds.
\end{lem}
\proof
For $n\in\N$,
\begin{eqnarray}
v_{a,d}(a)&=&v_{a,d}(a-1/n)\E_a\left[ e^{-q\tau_{a-1/n}^-}, \tau_{a-1/n}^-<\tau_{a+1/n}^+\right]\nonumber\\&&
+v_{a,d}(a+1/n)\E_a\left[ e^{-q\tau_{a+1/n}^+}, \tau_{a+1/n}^+<\tau_{a-1/n}^-\right]
+{\rm o}\left(\frac{1}{n}\right),\label{wyjasnienie}\end{eqnarray}
where the last term is bounded above by $\frac{1}{n}\P(\tau^+_{1/n}>d)$.
Moreover, by (\ref{twoex}), $$\E_a\left[ e^{-q\tau_{a+1/n}^+}, \tau_{a+1/n}^+<\tau_{a-1/n}^-\right]
=\frac{W^{(q)}(1/n)}{W^{(q)}(2/n)}$$
and by (\ref{two-sided-down}),
$$\E_a\left[ e^{-q\tau_{a-1/n}^-}, \tau_{a-1/n}^-<\tau_{a+1/n}^+\right]
=Z^{(q)}(1/n)-Z^{(q)}(2/n)\frac{W^{(q)}(1/n)}{W^{(q)}(2/n)}.
$$
Multiplying both sides of (\ref{wyjasnienie}) by two, subtracting $v_{a,d}(a-1/n)+v_{a,d}(a)$, dividing by $1/n$
gives:
\begin{eqnarray}
\frac{v_{a,d}(a)-v_{a,d}(a-1/n)}{1/n}&=&\frac{v_{a,d}(a+1/n)-v_{a,d}(a)}{1/n}\label{trudnosci0}\\&& + \frac{v_{a,d}(a+1/n)-v_{a,d}(a-1/n)}{1/n}\left[\frac{2W^{(q)}(1/n)}{W^{(q)}(2/n)}-1\right]\label{trudnosci1}
\\&&+v_{a,d}(a-1/n)2q\frac{\int_0^{1/n}W^{(q)}(y)\,dy-\frac{W^{(q)}(1/n)}{W^{(q)}(2/n)}\int_0^{2/n}W^{(q)}(y)\,dy}{1/n}\nonumber\\&&\label{trudnosci2b}
\end{eqnarray}
where we use (\ref{Zq}).
Now, since $W^{(q)}(0)=0$ and $W^{(q)\prime}(0)=\frac{2}{\sigma^2}$, we have:
$$\lim_{n\to\infty}\frac{W^{(q)}(1/n)}{W^{(q)}(2/n)}=\frac{W^{(q)\prime}(0)}{2W^{(q)\prime}(0)}=\frac{1}{2}.$$
Hence increment (\ref{trudnosci2b}) converges to
$v_{a,d}(a)2q(W^{(q)}(0)-\frac{1}{4}W^{(q)}(0))=0$ as $n\to\infty$.
Moreover,
\begin{eqnarray*}\lim_{n\to\infty}n\left[\frac{2W^{(q)}(1/n)}{W^{(q)}(2/n)}-1\right]&=&\lim_{n\to\infty}
-\frac{1}{2}\frac{2/n}{W^{(q)}(2/n)}\frac{W^{(q)}(2/n)-2W^{(q)}(1/n)}{1/n^2}\\&=&-\frac{1}{2}\frac{1}{W^{(q)\prime}(0)}W^{(q)\prime\prime}(0)<\infty
\end{eqnarray*}
and $\lim_{n\to\infty}(v_{a,d}(a+1/n)-v_{a,d}(a-1/n))=0$ by the continuity of the value function.
Hence increment (\ref{trudnosci1}) also tends to $0$ as $n\to\infty$.
Taking limit as $n\to\infty$ in (\ref{trudnosci0})-(\ref{trudnosci2b}) completes
the proof of (\ref{smoothpaste}).
\exit

All results of this section could be summarized in the following theorem.

\begin{Thm}\label{ThmMain2}
The value function $v_{a,d}(x)$ corresponding to the strategy $\pi_{a,d}$ is given in (\ref{basicid1}) - (\ref{smoothpaste}).
\end{Thm}

\sec{Examples}\label{sec:examples}
\subsection{Classic risk process with exponential jumps}

Assume that $X_t$ is Cram\'er-Lundberg risk process (\ref{CL}) with exponential claims $F(dz)=\xi e^{-\xi z}\,dz$
and intensity of their arrival $\lambda$.
Then under $\P^{\Phi(q)}$, where
$$\Phi(q)=\frac{{q} + \lambda -\xi c +\sqrt{({q} + \lambda-\xi
c)^2 + 4 c {q}\xi}}{2 c},$$
process $X_t$ is again Cram\'er-Lundberg risk process (\ref{CL}) with exponential claims
with parameter $\xi_q=\xi+\Phi(q)$ and intensity of Poisson arrival equal to $\lambda_q=\lambda\xi/\xi_q$ (see (\ref{tildepsi})).

\subsubsection{Dividend strategy $\pi^{a,\zeta}$ }

We start form the strategy $\pi^{a,\zeta}$, where we pay dividend until the Parisian ruin time.
The value function for the barrier strategy is given in (\ref{vax}) with
\begin{eqnarray}\label{expo}
 V^{(q)}(x)= e^{\Phi(q)x}\left(1-e^{-\left(\frac{c\xi_q-\lambda_q}{c}\right)x}\left( \frac{\lambda_q D_q}{c\xi_q -\lambda_q(1-D_q) }\right)\right),
\end{eqnarray}
where $$D_q=1-\int_0^\zeta \sqrt{\frac{c\xi_q}{\lambda_q}}e^{-(\lambda_q+c\xi_q)t}t^{-1}I_1(2t\sqrt{c\lambda_q\xi})dt$$
and $I_1(x)$ is modified Bessel function of the first kind
(see \cite{iczp} and \cite{{DassWu1}} for details).
Then solving (\ref{ceq}) we derive the optimal barrier:
$$a^{*}=\frac{c}{c\xi_q-\lambda_q}\log \left[\left( \frac{\lambda_q D_q}{c\xi_q -\lambda_q(1-D_q) }\right)\left(1-\frac{c\xi_q-\lambda_q}{c\Phi(q)}\right)^2\right].$$
From Corollary \ref {cor:explopt} it follows that barrier strategy with barrier $a^*$ is optimal strategy.

We present now the numerical analysis. 
We study Parisian dividends for different Parisian delays (Table 1): $\zeta=$ $0.1$, $0.3$, $0.7$, $2$ and for different initial capitals (Table 2):  $x=$ $2$, $5$, $10$, $50$. Moreover in Table 3 we compare the initial capitals when dividend payments are made till classical and Parisian ruin time.
In the first case we denote the value function by $v$ (see \cite{APP} for details concerning identifying it).
We choose the following parameters: $\xi=2$, $\lambda=2$, $c=2.5$ (see also Albrecher and Thonhauser \cite{AT08}).
\vspace{0.5cm}

\begin{center}
\begin{tabular}{|c|c|c|c|c|c|}
\hline
$\zeta$ & 0.1 & 0.3 & 0.7& 2  \\
\hline\hline
$a^{*}$
&3.54	&3.09	&2.40	&0.84\\
\hline
\end{tabular}
\vspace{0.5cm}

Table 1: The optimal barrier for different Parisian delays.
\end{center}

\begin{center}
\begin{tabular}{|c|c|c|c|c|c|c|}
\hline
x & 2 & 5 & 10& 50  \\
\hline\hline
$v(x)$ &12.57	& 15.71	& 20.71	& 60.71\\
\hline
$v^{a^*,\zeta}(x)$ &
13.38	&16.40	&21.40	&61.40\\
\hline
\end{tabular}
\vspace{0.5cm}

Table 2: The value function for classical and Parisian dividend problem for different initial capitals, when Parisian delay equals $\zeta=0.3$.
\end{center}

\begin{center}
\begin{tabular}{|c|c|c|c|c|c|c|}
\hline
x & 2 & 5 & 10& 50  \\
\hline
$x_1$ & 2.69 & 5.69 & 10.69& 50.69  \\
\hline
$v(x_1)=v^{a^*,\zeta}(x)$ &
13.38	&16.40	&21.40	&61.40\\
\hline
\end{tabular}
\vspace{0.5cm}

Table 3: The same value function for classical and Parisian dividend problem for different initial capitals, when Parisian delay equals $\zeta=0.3$.
\end{center}


\subsubsection{Dividend strategy $\pi_{a,d}$ }

In the second scenario $\pi_{a,d}$ we have delay between decision to pay and its implementation. Then the value function is given in (\ref{basicid1}) with
the scale function $W^{(q)}$ given by:
$$
W^{(q)}(x) = c^{-1}\left(A_+ \te{q^+(q)x} - A_-
\te{q^-(q)x}\right),
$$
where $A_\pm = \frac{\xi + q^\pm(q)}{q^+(q)-q^-(q)}$ with
$q^+(q)=\F(q)$ and $q^-(q)$ being the smallest root of $\psi(\th)=q$:
$$q^-(q)=\frac{{q} + \lambda -\xi c - \sqrt{({q} + \lambda-\xi
c)^2 + 4 c {q}\xi}}{2 c}.$$
Moreover, to identify $v_{a,d}(a)$ note that, by (\ref{basicid2}), (\ref{onesideddwon}) and lack of memory of exponential distribution,
for $x\geq a$ we have,
\begin{eqnarray}
v_{a,d}(x)&=&\E_{x-a}e^{-q d}\left[X_d, \tau_0^->d\right]+v_{a,d}(a)e^{-q d}\P_{x-a}\left(\tau_0^->d\right)\nonumber\\&&+v_{a,d}(a)\E_{x-a}\left[e^{-q \tau^-_0}, \tau_0^-\leq d\right]
\int_{0}^a\frac{W^{(q)}(a-y)}{W^{(q)}(a)}\xi e^{-\xi y}\,dy\nonumber\\&=&
e^{-qd}\int_0^\infty y\P_{x-a}(\tau_0^->d, X_d\in dy)\nonumber\\
&&+v_{a,d}(a)e^{-qd}\P_{x-a}\left(\tau_0^->d\right)\nonumber\\&&+
v_{a,d}(a)\int_0^d e^{-qt}\P_{x-a}(\tau_0^-\in dt)\int_{0}^a\frac{W^{(q)}(a-y)}{W^{(q)}(a)}\xi e^{-\xi y}\,dy. \label{basicid22}
\end{eqnarray}
Further,
$$\P_{z}(\tau_0^->d, X_d\in dy)=\P_z(X_d\in dy)-\int_0^d\P_z(\tau_0^-\in dt)\int_0^\infty\P_{-w}(X_{d-t}\in dy)\xi e^{-\xi w}\,dw$$
with $\P_z(X_s\in dy)=e^{-\lambda s}\delta_{cs}(dy)+\sum_{k=1}^\infty\frac{(\lambda s)^k}{k!}e^{-\lambda s}\frac{\xi^k}{(k-1)!}(cs-y)^{k-1}e^{-\xi (cs-y)}\,dy$.
Finally, by \cite[Prop. IV.1.3, p. 101]{Assbook}:
$$
\P_x(\tau_0^-<t)=\lambda e^{-(1-\lambda/\xi)x}-\frac{1}{\pi}\int_0^{\pi} \frac{f_1(\theta)f_2(\theta)}{f_3(\theta)}\,d\theta,
$$
where
\begin{eqnarray*}
f_1(\theta)&=&\frac{\lambda}{\xi} \exp\left\{ 2\sqrt{\frac{\lambda}{\xi}}t \cos \theta -(1+\lambda/\xi)t +x(\sqrt{\frac{\lambda}{\xi}}\cos \theta -1)\right\},\\
f_2(\theta)&=&\cos\left(x\sqrt{\frac{\lambda}{\xi}}\sin \theta\right)-\cos\left(x\sqrt{\frac{\lambda}{\xi}}\sin \theta+2\theta\right),\\
f_3(\theta)&=&1+\frac{\lambda}{\xi}-2\sqrt{\frac{\lambda}{\xi}}\cos \theta.\\
\end{eqnarray*}

\subsection{Brownian motion with drift - small claims}

Assume that
$$X_t=\sigma B_t+ct,$$
where $\sigma,c >0$ and $B_t$ is a standard Brownian motion.
Then under $\P^{\Phi(q)}$, where
$\Phi(q)=\frac{c_q-c}{\sigma^2}$ with $c_q=\sqrt{c^2+2q\sigma^2}$, we have by (\ref{tildepsi}) that $X_t= \sigma B_t+c_qt$.

\subsubsection{Dividend strategy $\pi^{a,\zeta}$ }
Considering strategy $\pi^{a,\zeta}$ with Parisian delay at ruin, the value function is given in (\ref{vax}) with
\begin{eqnarray}\label{expo2}
 V^{(q)}(x)=e^{\Phi(q)x}\left(1- e^{-(2c_q\sigma^{-2})x} \frac{\Psi\left(\frac{c_q}{\sigma}\sqrt{\frac{\zeta}{2}}\right)-\frac{c_q}{\sigma}\sqrt{\frac{\zeta\pi}{2}}}{\Psi\left(\frac{c_q}{\sigma}\sqrt{\frac{\zeta}{2}}\right)+\frac{c_q}{\sigma}\sqrt{\frac{\zeta \pi}{2}}}\right),
\end{eqnarray}
where $$\Psi(x)=2\sqrt{\pi}x\mathcal{N}(\sqrt{2}x)-\sqrt{\pi}x+e^{-x^2}$$
and $\mathcal{N}$(.) is a cumulative distribution function for the standard normal distribution (for details see \cite{iczp} and \cite{{DassWu2}}).
Hence by (\ref{ceq}) the optimal barrier is given by:
$$a^{*}=\frac{\sigma^2}{2c_q}\log\left[\frac{\Psi\left(\frac{c_q}{\sigma}\sqrt{\frac{\zeta}{2}}\right)-\frac{c_q}{\sigma}\sqrt{\frac{\zeta\pi}{2}}}{\Psi\left(\frac{c_q}{\sigma}\sqrt{\frac{\zeta}{2}}\right)+\frac{c_q}{\sigma}\sqrt{\frac{\zeta \pi}{2}}}\left(1-\frac{2c_q}{c_q-c}\right)^2\right]$$
and by Corollary \ref{wniosek} it is the optimal strategy.

Similarly like for the Cram\'er-Lundberg process we present here the numerical analysis. We study Parisian dividends for different Parisian delays (Table 4): $\zeta=$ $0.1$, $0.3$, $0.7$, $2$ and for different initial capitals (Table 5):  $x=$ $2$, $5$, $10$, $50$. In Table 6 we compare initial capitals for classical and Parisian dividend problems.
We take $c=2.5$ and $\sigma=2$. 
\vspace{0.5cm}

\begin{center}
\begin{tabular}{|c|c|c|c|c|c|}
\hline
$\zeta$ & 0.1 & 0.3 & 0.7& 2  \\
\hline\hline
$a^{*}$
&4.48	&3.89	&3.12	&1.17\\
\hline
\end{tabular}
\vspace{0.5cm}

Table 4: The optimal barrier for different Parisian delays.
\end{center}

\begin{center}
\begin{tabular}{|c|c|c|c|c|c|c|}
\hline
x & 2 & 5 & 10& 50  \\
\hline\hline
$v(x)$ &
20.49	& 24.72	& 29.72	& 69.72\\
\hline
$v^{a^*,\zeta}(x)$ &
23.00	&26.11	&31.11	&71.11\\
\hline
\end{tabular}
\vspace{0.5cm}

Table 5: The value function for classical and Parisian dividend problem, when Parisian delay equals $\zeta=0.3$.
\end{center}

\begin{center}
\begin{tabular}{|c|c|c|c|c|c|c|}
\hline
x & 2 & 5 & 10& 50  \\
\hline
$x_1$ & 3.40 & 6.39 & 11.39& 51.39  \\
\hline
$v(x_1)=v^{a^*,\zeta}(x)$ &
23.00	&26.11	&31.11	&71.11\\
\hline
\end{tabular}
\vspace{0.5cm}

Table 6: The same value function for classical and Parisian dividend problem, when Parisian delay equals $\zeta=0.3$.
\end{center}

\subsubsection{Dividend strategy $\pi_{a,d}$ }
To get value function $v(x)$ given in (\ref{basicid1}) for the case $\pi_{a,d}$ with Parisian delay at the moment of payment of dividends, we
have to identify first the scale function:
$$
W^{(q)}(x) = \frac{1}{\s^2\d} \left(\te{(-\omega+\d) x} -
\te{-(\omega+\d)x}\right),
$$
where $\d = \s^{-2}\sqrt{c^2 + 2q\s^2}$ and $\omega = c/\s^2$.
Then, similarly like in the previous example,
\begin{eqnarray}
v_{a,d}(x)&=&
e^{-qd}\int_0^\infty y\P_{x-a}(\tau_0^->d, X_d\in dy)\nonumber\\
&&+v_{a,d}(a)\left[e^{-qd}\P_{x-a}\left(\tau_0^->d\right)+\E_{x-a}\left[e^{-q\tau_0^-},\tau_0^-\leq d\right]\right]\label{basicid23}
\end{eqnarray}
and $$\P_{z}(\tau_0^->d, X_d\in dy)=\P_z(X_d\in dy)-\int_0^d\P_z(\tau_0^-\in dt)\P(X_{d-t}\in dy)$$
with $X_s$ having $N(cs,s)$ distribution and $\tau_0^-$ having inverse Gaussian distribution:
$$\P_x(\tau_0^-\in dt)=\frac{x}{\sqrt{2\pi t^3}}e^{-xc}e^{-\frac{1}{2}\left(x^2t^{-1}+c^2t\right)}\,dx.$$

\section*{Acknowledgements}
This work is partially supported by the Ministry of Science and
Higher Education of Poland under the grants N N201 394137
(2009-2011) and N N201 525638 (2010-2011).

\end{document}